\documentclass[epsfig,12pt]{article}
\usepackage{epsfig}
\usepackage{amsfonts}
\begin {document}

\title {EVENT HORIZON OF THE MONOPOLE-QUADRUPOLE SOLUTION: GEOMETRIC AND THERMODYNAMIC PROPERTIES}

\author{J.L. Hern\'andez-Pastora\thanks{E.T.S. Ingenier\'\i a
Industrial de B\'ejar. Phone: +34 923 408080 Ext
2263. Also at +34 923 294400 Ext 1527. e-mail address: jlhp@usal.es} $^{1}$ and L.Herrera\thanks{Also at U.C.V. Venezuela. email: lherrera@usal.es} $^2$ \\
\\
$^1$ Departamento de Matem\'atica Aplicada \\
and Instituto Universitario de F\'\i sica Fundamental y Matem\'aticas. \\ Universidad de Salamanca.  Salamanca,
Spain.  \\
\\
$^2$Departamento   de F\'{\i}sica Te\'orica e Historia de la  Ciencia,\\
Universidad del Pa\'{\i}s Vasco, Bilbao, Spain}

\date{\today} \maketitle

\begin{abstract}
We investigate the general geometric properties of the surface of infinite red-shift corresponding to the event horizon of the static and axisymmetric  solution of the Einstein vacuum equations that only possesses mass $M$ and quadrupole moment $Q$. The deformation of the Schwarzschild surface $r=2M$ produced by the quadrupole moment is shown, and the range of values of this multipole moment is specified, which    preserves a regular, closed, continuous and differentiable surface. Some thermodynamic consequences and speculations ensuing from our results are discussed.
\end{abstract}


\newpage

\section{Introduction}
As it is well known \cite{1},
the only static and asymptotically-flat vacuum space-time possessing
a regular horizon is the Schwarzschild solution. All the other Weyl
exterior solutions \cite{2}--\cite{2IV}, exhibit singularities in the physical
components of the Riemann tensor  at the horizon.

For not particularly intense gravitational fields and small
fluctuations,  deviations from spherical
symmetry may be described as perturbations of the Schwarzschild exact solution \cite{Letelier}.

However, such perturbative scheme will eventually fail in regions
close to the horizon (although strictly speaking the term
``horizon'' refers to the spherically symmetric case, we shall use
it when considering the $r=2M$ surface, in the case of small
deviations from sphericity). Indeed,  as we approach the horizon,
any finite perturbation of the Schwarzschild space-time becomes
fundamentally different from the corresponding exact solution
representing the quasi--spherical space-time,  even if the latter is
characterized by parameters whose values are arbitrarily close to
those corresponding to Schwarzschild metric  \cite{4}--\cite{4IV}. This, of
course,  is just an expression of the Israel theorem  (for observational differences between black holes and naked singularities see \cite{VE}, \cite{VK} and references therein).

Therefore, for strong gravitational fields, no matter how small the
multipole moments of the source are (those higher than monopole),
there exists a bifurcation between the perturbed Schwarzschild
metric and all the other Weyl metrics (in the case of gravitational
perturbations).

Examples of such a  bifurcation have been brought out for  the $\gamma$ metric
\cite{zipo}--\cite{zipoVII} and   for the space-time possessing only monopole and quadrupole moments (hereafter  M--Q space-time \cite{yo}, \cite{yobis}, \cite{yoE}), in  the study
of the trajectories of test particles for
orbits close to $r=2M$ \cite{HS}, \cite{herreramq}.

The influence of the quadrupole moment on the motion of test
particles within the context of Erez--Rosen metric \cite{erroz} has
been investigated by many authors (see \cite{quevedo1}--\cite{quevedo4} and
references therein).

The purpose of this paper is to study further the properties of the  M--Q space-time in order to bring out the bifurcation mentioned above, and to contrast our results with those obtained for other Weyl metrics.

The
rationale for the choice of  the  M--Q space-time is based on the fact  that   its relativistic multipole
structure  (particularly that of  a sub--class of this
solution M--Q$^{(1)}$ \cite{yo}, which, being an exact solution, corresponds to the first order in quadrupole moment in M--Q) may be interpreted as a quadrupole
correction to the Schwarzschild space-time, and therefore
represents a good candidate among known Weyl solutions, to describe
small deviations from spherical symmetry.

In this paper we shall see that  unlike in  the $\gamma$--metric   \cite{zipoVII}, the area does not vanish at the horizon. However, the surface gravity  diverges  as is the case for the $\gamma$--metric \cite{ita}. We shall calculate explicitly the surface of infinite red--shift corresponding  of the M--Q space-time and determine the range of values of the quadrupole moment for which such a surface is close, continuous and differentiable. Next we shall calculate the surface gravity  which exhibits a singularity along the symmetry axis as we approach the event horizon. Some thermodynamic consequences derived from these results are discussed.

\section{The M-Q solution}

In \cite{msa} a so-called Multipole Symmetry Adapted  coordinate system  ($MSA$)  was defined which  allows us to write the $g_{tt}$ metric component of  pure multipole solutions in General Relativity, in a form  resembling the classical multipole potential. The existence of such a type of coordinate system was proved in \cite{msa}. The relation between the $MSA$ system of coordinates $(r,y\equiv \cos\theta)$ and the Weyl standard coordinates $(R,\omega\equiv\cos\Theta)$ is given by the following expression:
\begin{eqnarray}
 r=R\left[1+\sum_{n=1}^{\infty}f_n(\omega)\frac{1}{R^n}\right] \nonumber \\
y=w+\sum_{n=1}^{\infty}g_n(\omega)\frac{1}{R^n} \ ,
\label{trans}
\end{eqnarray}
$f_n(\omega)$, $g_n(\omega)$ being  polynomials in the angular variable $\omega$ explicitly calculated in \cite{msa}, where the consistency of this expansion, and the convergence of the series is discussed. In particular, for the case of the solution with a finite number of multipole moments given by the monopole and the quadrupole moments, the corresponding $MSA$ coordinates can be written as follows:
\begin{eqnarray}
r&=&r_{s}\left[1-q P_2(y_{s}) \lambda_s^3-\frac{q}{28}(35y_{s}^4+6y_{s}^2-9)\lambda_s^4-\right.\nonumber\\
&+&\left.\lambda_s^5\left(\frac{q}{28}(105y_{s}^4-66y_{s}^2+1)-q^2(12y_{s}^4-9y_{s}^2+1)\right)+
\dots\right]\nonumber\\
y&=&y_{s}\left[1-(1-y_{s}^2)\left(q\lambda_s^3+\frac{5q}{28}(7y_{s}^2+9)\lambda_s^4+\right.\right.\nonumber\\
&+&\left.\left.\frac{q}{35}\lambda_s^5\left((175y_{s}^2+57)+
q(126-336y_{s}^2)\right)+\dots \right)\right]
\end{eqnarray}
where $P_n(y)$ stands for the Legendre polynomials, $r_{s}$, and $y_{s}$ denote the standard Schwarzschild coordinates, $\lambda_s\equiv M/r_{s}$, and $q\equiv Q/M^3$, ($Q$ stands for the quadrupole moment).

The relevance and the physical and mathematical interest of these systems of coordinates become evident   from the fact that  they are related with the existence of certain symmetries of the static and axisymmetric vacuum field equations, which generalize the action of those  symmetry groups  on the classical equations \cite{cqg}. For the monopole solution (spherical symmetry) the $MSA$ system of coordinates is exactly the standard Schwarzschild coordinate, and the associated symmetry group of the field equations allow us to determine univocally the system of coordinates by means of solving certain  Cauchy problem.
In a recent work \cite{phrd} the $MSA$ system of coordinates has been used to calculate relativistic corrections of the perihelion advance for orbital test particles.

The M--Q solution of the static  and axisymmetric vacuum equations (see \cite{phrd}, \cite{yo}) can be written in the corresponding $MSA$ system of coordinates ${\hat x}\equiv(r,y)$ as follows
\begin{eqnarray}
g_{tt}(\hat x)&=&-1+2 \left[  \hat\lambda+\frac{Q}{M^3}\hat\lambda^3 P_{2}(y)\right] \nonumber\\
 g_{rr}(\hat x)&=&\frac{1}{1-2\hat \lambda}\left[1+(1-2\hat\lambda)\sum_{i=3}^{\infty}\hat\lambda^i
U_i(y,Q)\right] \nonumber\\
g_{yy}(\hat x)&=&\frac{1}{1-y^2}\frac{M^2}{\hat \lambda^2}\left[1+\sum_{i=3}^{\infty}\hat\lambda^i
D_i(y,Q)\right] \nonumber\\
g_{\varphi \varphi}(\hat x)&=&(1-y^2)\frac{M^2}{\hat \lambda^2}\left[1+\sum_{i=3}^{\infty}\hat\lambda^i
T_i(y,Q)\right] \ ,
 \label{chachis}
\end{eqnarray}
where $\hat\lambda\equiv M/r$, and  $U_i$, $D_i$ and $T_i$ denote polynomials in the angular variable $y$ of even order depending on the quadrupole moment $Q$. The explicit form of some polynomials are given in \cite{phrd}.

\section{The event horizon}

In this work we want to make use of the $MSA$ system of coordinates to calculate the event horizon of the exact monopole-quadrupole solution of the static and axisymmetric vacuum Einstein equations.

In $MSA$ coordinates   the $g_{tt}$ metric component of the M--Q solution takes the form
\begin{equation}
g_{00}=-1+2 V(r,y) ,\label{g00}
\end{equation}
with $\displaystyle{V(r,y)=\frac Mr+\frac{Q}{r^3}P_2(y)}$, which is a reminiscence of the classical multipole potential but now constructed with the relativistic monopole moment $M$ and the quadrupole moment $Q$ respectively. The event horizon of this static solution is given by the equipotential surface $V(r,y)=\beta$ with $\beta=1/2$. Equivalently,  the surface of infinite red-shift defined by the condition $g_{tt}\equiv \xi^{\alpha}\xi_{\alpha}=0$ ($\xi^{\alpha}$ being the time-like Killing vector) can be calculated by solving the following algebraic equation
\begin{equation}
r^3-2 M r^2-2 Q P_2(y)=0.
\label{algebraic}
\end{equation}
Therefore, the appropriate root of this equation is a function $r=r(y)$ which represents the generatrix curve of  revolution surface describing the event horizon of the axisymmetric solution. Let us note that this curve is symmetric with respect to the equatorial  plane due to the symmetry of the solution and therefore, the study of $r=r(y)$ can be restricted to the value of $y \in (0,1)$, or equivalently the angular coordinate $\theta \in (0,\pi/2)$.

In order to determine the real roots of this third degree algebraic equation we must define the different domains of the equation in terms of the values for $y=\cos\theta$. It depends on the value of the quantity
\begin{equation}
 d = q P_2(y) \left(q P_2(y)+\frac{16}{27}\right) \equiv a(a+2 \alpha) ,\label{ddea}
\end{equation}
with the notation $a\equiv q P_2(y)$, $\alpha\equiv 8/27$, and $q\equiv Q/M^3$, as follows:
\begin{eqnarray}
DI &:&d<0 , a \in (-2 \alpha, 0),\nonumber\\
DII &:&d>0 , a \in (-\infty, -2\alpha) \cup (0, \infty), \nonumber\\
DIII &:&d=0 , a=-2 \alpha, a=0.
\label{domains}
\end{eqnarray}

The region defined by $DI$ leads to  three real roots of the algebraic equation (\ref{algebraic}), as well as the region $DIII$ (where at least two of these roots are equals), whereas region $DII$ provides only one real root.

Let us first analyze the region $DIII$:  the solutions of the algebraic equation for this region are the following
\begin{eqnarray}
 r_{d=0}(y)\equiv r_1(y)&=&2 M\left[\frac 13+(a+\alpha)^{1/3}\right], \nonumber\\
r_{d=0}(y)\equiv r_{2,3}(y)&=&2 M\left[\frac 13-\frac 12(a+\alpha)^{1/3}\right].
\label{digualq}
\end{eqnarray}
It is easy to show that we have positive real roots leading to a good matching with the region $DII$ only if $a=0$, which corresponds to the angular value $y_1^0=1/\sqrt{3}$: the other possible point of this region is $y_2^0=y_1^0\sqrt{1-4\alpha/q}$, which corresponds to the condition $a=-2\alpha$, as can be seen from Figure 1. However, the roots for that value are:
\begin{equation}
  r_1(y_2^0)= -\frac{2M}{3} , \qquad  r_{2,3}(y_2^0)= \frac{4M}{3},
\end{equation}
and they only match with the unique real root for the region $DII$ ($d>0$) providing a negative value of the curve:
\begin{equation}
 r_{d>0}(y)=M\left[(a+\alpha+\sqrt{d})^{1/3}+(a+\alpha-\sqrt{d})^{1/3}+\frac 23\right] ,
r_{d>0}(y_2^0)=-\frac{2M}{3}.
\label{dmayorq}
\end{equation}

\begin{figure}[h]
\caption{}
\begin{minipage}{.49\linewidth}
\centering
\includegraphics[height=2.8in,width=2.8in]{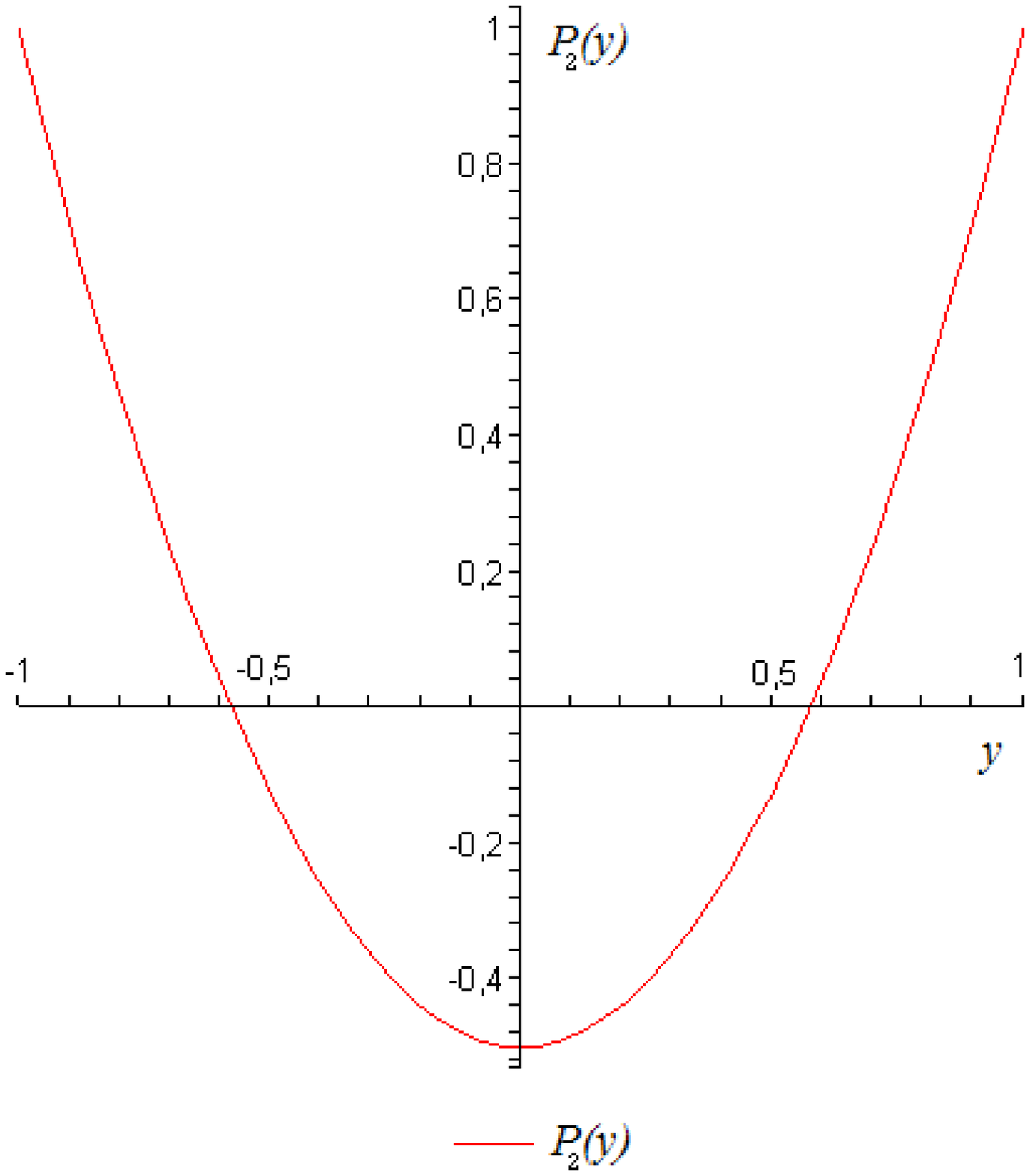}
{Legendre polynomial of degree 2 in the variable $y\equiv \cos\theta$.}
\end{minipage}
\begin{minipage}{.49\linewidth}
\centering
\includegraphics[height=2.5in,width=2.5in]{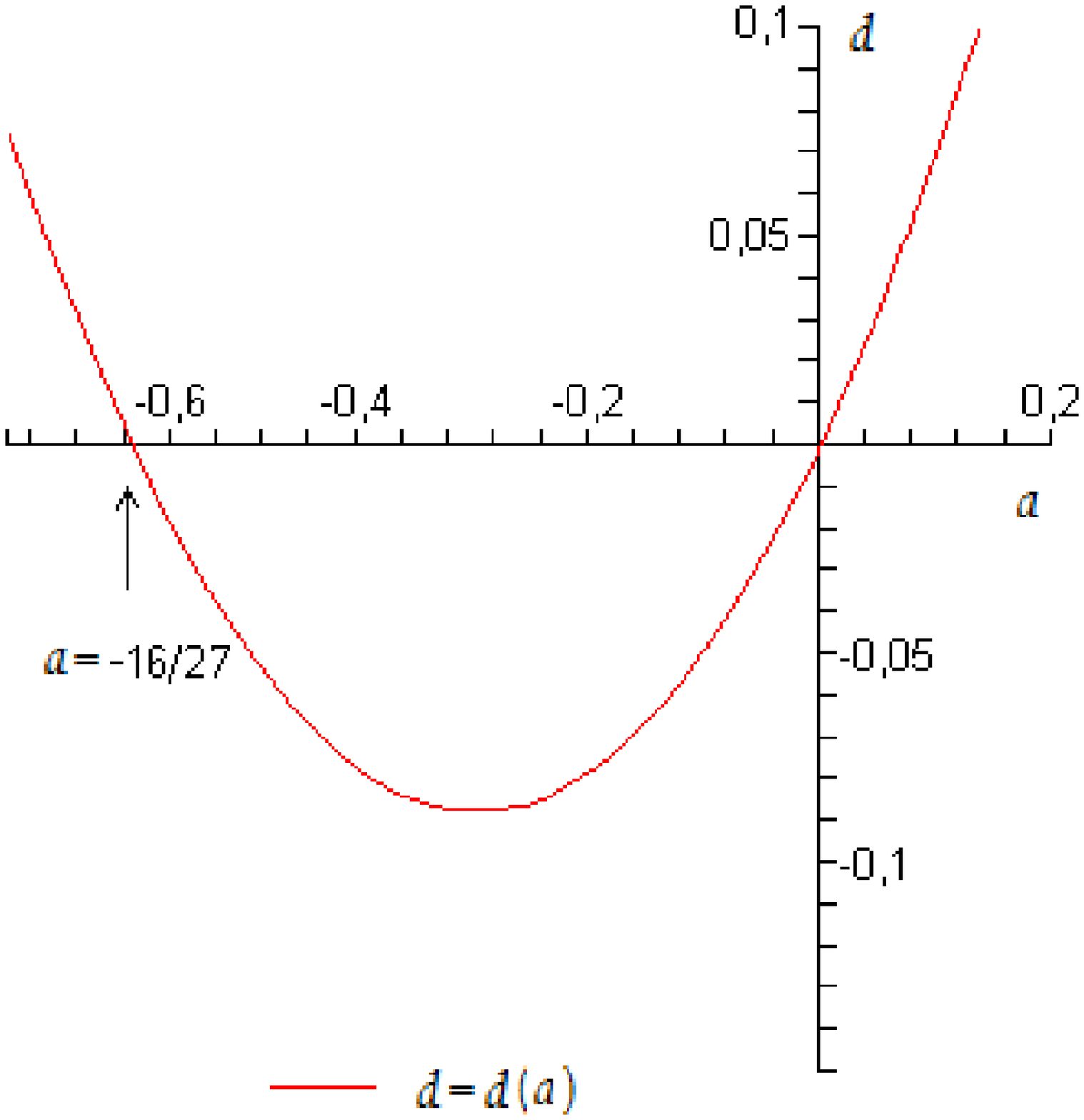}
{The function $d=d(a)$  in terms of $a$ (\ref{ddea}). The two points where $d$ vanishes are $a=0$ and $a=-2\alpha$ ($\alpha\equiv 8/27$).}
\end{minipage}
%\caption{}
\label{fig1}
\end{figure}

Therefore we must limit the value of the parameter $q$ within the range $-2\alpha < q \leq 4 \alpha$, where the value of $y_2^0$ is not real, to preserve the continuity of the curve $r=r(y)$.

Next, as indicated in  (\ref{domains}) , the region $DII$ ($d>0$) splits into two different open intervals given by  the conditions $a>0$ and $a<-2\alpha$. The condition $a>0$ leads to $y<y_1^0$ (for $q<0$) or $y>y_1^0$ (for $q>0$) whereas the condition $a<-2\alpha$ cannot be fulfilled if the regularity range of the parameter $q \in (-2\alpha,4\alpha]$  holds.

Hence, we can conclude that the different domains for the evaluation of the real roots of (\ref{algebraic}) are given by
\begin{equation}
q \in (0,4 \alpha]
\left\{
\begin{array}{c|c}
d>0& y>y_1^0\\
d=0& y=y_1^0 \\
d<0& y<y_1^0
\end{array}
\right.
\end{equation}
\begin{equation}
q \in (-2 \alpha, 0)
\left\{
\begin{array}{c|c}
d>0& y<y_1^0\\
d=0& y=y_1^0 \\
d<0& y>y_1^0
\end{array}
\right.
\end{equation}

Finally,  the region $DI$ ($d<0$)  has the following three real roots:
\begin{eqnarray}
r(y)&=&\frac 23 M\left(2 \cos \frac{\kappa}{3}+1\right)\equiv r_{d<0},\nonumber\\
r(y)&=&\frac 23 M\left(2 \cos \frac{\kappa}{3}\pm \sqrt{3} \sin \frac{\kappa}{3}+1\right)\equiv r_{\pm},
\label{dmenorq}
\end{eqnarray}
where $\kappa$ is defined by $\displaystyle{\cos \kappa=1+\frac{a}{\alpha}}$. Let us note that these solutions $r_{\pm}$, $r_{d<0}$ are well defined only if  $-2\alpha<a<0$ which in fact agrees with the range of $a$ for this region (\ref{domains}). Additionally $\displaystyle{\kappa=\arccos\left(1+\frac{a}{\alpha}\right)}$  exhibits a good behavior since it is uniquely valued at the boundary point $y=y_1^0$, i.e. $\kappa$ has a single value ($\kappa=0$) at the limit $a\rightarrow0$ ($y=y_1^0$) since $\kappa \in [0,\pi]$ where $-2\alpha<a<0$.

The curve $r=r(y)$ matches appropriately at the boundary between the regions $DII$ and $DIII$ where
\begin{equation}
 r_{d>0}(y=y_1^0)= 2 M = r_{1}(y=y_1^0)
\end{equation}
and the curve is completely continuous for all  range $0\leq y \leq 1$ if we consider the curve $r=r(y)$ to be  equal to $r_{d<0}$ (\ref{dmenorq}) at the region $DI$ ($d<0$), the other  roots $r_{\pm}$ being neglected.

In conclusion,  M--Q solution possesses an event horizon defined by a generatrix curve which is continuous for the whole range $0\leq \theta \leq \pi$ as follows:
\begin{equation}
q \in (0,4 \alpha] : \qquad r(y)=
\left\{\nonumber
\begin{array}{cc}
r_{d<0} \quad , & \quad 0\leq y \leq y_1^0 \nonumber\\
r_{d>0} \quad , & \quad y_1^0\leq y \leq 1 \nonumber
\end{array}
\right.\nonumber
\nonumber
\label{generatrix1}
\end{equation}
\begin{equation}
q \in (-2 \alpha, 0) : \qquad r(y)=
\left\{
\begin{array}{cc}
r_{d>0} \quad , & \quad 0\leq y \leq y_1^0\\
r_{d<0} \quad , & \quad y_1^0\leq y \leq 1
\end{array}
\right.
\label{generatrix2}
\end{equation}
 Let us note that the continuous curve $r=r(y)$ must be a piecewise-defined function  since  $\kappa$ is not
defined for the region $d>0$ $DII$ where $a>0$. Nevertheless it can be shown that  functions $r_{d<0}$ and $r_{d>0}$ are equivalent since we can rewrite $r_{d<0}$ (\ref{dmenorq}) by taking into account that
\begin{equation}
 \cos \kappa=1+\frac{a}{\alpha} \Leftrightarrow \cos\frac{\kappa}{3}=\frac{1}{2
\alpha^{1/3}}\left[(a+\alpha+\sqrt{d})^{1/3}+(a+\alpha-\sqrt{d})^{1/3}\right]
\end{equation}
and therefore
\begin{equation}
 r_{d<0}(y)=M\left[(a+\alpha+\sqrt{d})^{1/3}+(a+\alpha-\sqrt{d})^{1/3}+\frac 23\right]
\end{equation}
which represents the root $r_{d>0}$.

Finally, we can prove that the surface of the event horizon defined by the generatrix curve (\ref{generatrix1})-(\ref{generatrix2}) is class $C^1$, i.e.  $r(y)$ is differentiable and the derivative is continuous. The derivative of the generatrix curve (\ref{generatrix1})-(\ref{generatrix2}) with respect to the angular parameter $\theta$ is given by the following expressions:
\begin{eqnarray}
\frac{d r_{d<0}}{d \theta}&=& -\sqrt{1-y^2}\frac{4M}{3}\frac{q y}{\sqrt{-d}}\sin \frac{\kappa}{3}\nonumber\\
\frac{d r_{d>0}}{d \theta}&=& -\sqrt{1-y^2}\frac{q M y}{\sqrt{d}} \left[(a+\alpha+\sqrt{d})^{1/3}-(a+\alpha-\sqrt{d})^{1/3}\right] \ ,
\label{derivadas}
\end{eqnarray}
and the continuity of the derivative at the boundary ($y=y_1^0$) is fulfilled:
\begin{eqnarray}
 \left(\frac{d r_{d<0}}{d y}\right)_{y=y_1^0}&=&\lim_{a\rightarrow 0}\frac{4M}{3}\frac{q y}{\sqrt{-d}}\sin
\frac{\kappa}{3}=\frac{3 q M}{2\sqrt{3}}\nonumber\\
 \left(\frac{d r_{d>0}}{d y}\right)_{y=y_1^0}&=&\lim_{a\rightarrow 0}\frac{q M y}{\sqrt{d}}
\left[(a+\alpha+\sqrt{d})^{1/3}-(a+\alpha-\sqrt{d})^{1/3}\right]=\frac{3qM}{2\sqrt{3}}.\nonumber\\
\end{eqnarray}
In addition, for any  value of the parameter $q$ within the range considered $(-2 \alpha,4\alpha]$, the derivative of the curve $r=r(y)$ is also continuous at the points intersecting  the symmetry axis $y=\pm 1$, as well as at the equatorial plane ($y=0$), since both  derivatives of $r_{d>0}$ and $r_{d<0}$ are zero at those points.

Hence, the surface of the event horizon is continuous and  differentiable  everywhere for the range  of values of the quadrupole moment $Q \in M^3 (-2 \alpha,4\alpha]$. Figures 2 to 4,  show different aspects of the generatrix curve as well as the surface of the event horizon.

It should be observed that although pathological behaviour might be expected for large (absolute) values of the  quadrupole moment, it is not clear why such a behaviour appears outside the range established above. We currently do not know  if there is any physical significance behind that range.

\section{The Area}

\vspace{2mm}

The area of the event-horizon surface $r=r(y)$ for the M--Q solution is given by the following expression:
\begin{equation}
A=\int_0^{2\pi}\int_{-1}^1\sqrt{\hat g_{yy}g_{\varphi\varphi}} dy d\varphi, \label{areadef}
\end{equation}
where
\begin{equation}
\hat g_{yy}=g_{rr}\left(\frac{dr(y)}{dy}\right)^2+g_{yy},
\end{equation}
and $g_{rr}$, $g_{yy}$, $g_{\varphi\varphi}$ are the metric components of the M--Q solution given by expressions (\ref{chachis}) in the corresponding MSA system of coordinates. As we have previously seen, the event horizon for the non-spherical case  depends on the angular variable $\theta$ ($\cos\theta=y$) for the range defined in (\ref{generatrix1})-(\ref{generatrix2}) and hence its derivative with respect to the variable $y$ is different from zero. Nevertheless, if we expand the expressions of the surface (\ref{generatrix1})-(\ref{generatrix2}) as well as its derivative (\ref{derivadas}) in terms of the quadrupole  parameter $q$, we obtain
\begin{eqnarray}
r(y)&\simeq& 2M \left(1-q^2\frac{P_2(y)^2}{8 \alpha}\right)+O(q^3)\nonumber\\
\left(\frac{dr(y)}{dy}\right)^2&\simeq&q^3\frac{M^2}{3\alpha^2}P_2(y) y^2+O(q^4),
\label{approx}
\end{eqnarray}
and we could approximate the event horizon to be equal to the Schwarzschild surface $r\simeq r_{s}=2M$ and $\displaystyle{\frac{d r}{dy}\simeq 0}$, since we are interested in the behaviour of the area for  very slight deviations from spherical symmetry. Equivalently we can say that $r(y)=2M+\epsilon$, where $\epsilon\equiv \displaystyle{\frac 43 M(-1+\cos \frac{\kappa}{3})}$ representing the difference between  the event horizon of the M--Q solution  and the corresponding surface in the spherical case, is a small quantity of order $q^2$ \footnote{It can be seen from figures 3--4 that $\displaystyle{\frac rM=2+p}$ where $p\equiv \epsilon/M$ with $-2/3\leq p\leq 1/3$, and $p=0$ for $q=0$ leading to $r(y)=2M$.}.

Therefore the area of the event horizon could be defined as follows:
\begin{equation}
A\simeq 2\pi \int_{-1}^1\sqrt{g_{yy}g_{\varphi\varphi}} dy,  \label{areaprox}
\end{equation}
 Since we want to know how different is the area of the event horizon for the non-spherical case for
slight  deviations  with respect to the Schwarzschild space-time, let us consider the M--Q$^{(1)}$ solution.

M--Q$^{(1)}$ is  a subclass of the   M--Q solution, that represents a small deformation of  Schwarzschild for a small parameter $q$. In fact, the M--Q$^{(1)}$ solution corresponds to  the first order in the parameter $q$ in an expansion of the solution M--Q in power series of that parameter.

 When written in prolate spheroidal coordinates $\{x,y\}$, the event horizon is defined by $x=1$
($r=r_{s}=2M$) if $q\leq 8/5$, and  metric functions of the solution are:
 \begin{eqnarray}
 e^{-2\Psi}&\simeq& \left(\frac{x+1}{x-1}\right)^A e^B+O(q^2),\nonumber\\
 e^{2 \gamma}&\simeq & \frac{x^2-1}{x^2-y^2} \left(\frac{x-1}{x+1}\right)^C e^D+O(q^2),
 \label{mq1}
 \end{eqnarray}
 where
 \begin{eqnarray}
 A&=& 1+\frac 54 q\left(-1+\frac{(3y^2-1)(3x^2-1)}{4}\right),\nonumber\\
 B&=& -\frac 58 qx\left(-\frac{4}{x^2-y^2}+3(3y^2-1)\right),\nonumber\\
 C&=& -\frac{15}{4} q x (1-y^2),\nonumber\\
 D&=& -\frac{15}{2} q(1-y^2)-\frac 52 q (x^2+y^2)\frac{1-y^2}{(x^2-y^2)^2},
 \label{mq1detal}
 \end{eqnarray}
 and the terms of order higher than $q$ have been neglected.

 The area of the event horizon is
 \begin{equation}
 A=\lim_{x\rightarrow 1} M^2\int_0^{2\pi}\int_{-1}^1 e^{-2\Psi}\sqrt{e^{2\gamma}(x^2-y^2)(x^2-1)} dy
d\varphi,
 \end{equation}
 or, using (\ref{mq1})-(\ref{mq1detal})
 \begin{equation}
 A=2\pi M^2 \lim_{x\rightarrow 1} \int_{-1}^1 (x+1)^{1-(C/2-A)}(x-1)^{1+(C/2-A)}e^{B+D/2} dy.
 \end{equation}

Then,  taking into account that
 \begin{equation}
 \lim_{x\rightarrow 1} (x+1)^{1-(C/2-A)}=4 \quad , \quad \lim_{x\rightarrow 1} (x-1)^{1+(C/2-A)}=1,
 \end{equation}
  we have that
 \begin{equation}
 A\simeq 2\pi 4M^2  \int_{-1}^1 e^{-\frac 58 q (1+3y^2)} dy,
 \end{equation}
 which  evaluated up to order $q$ produces
   \begin{equation}
 A\simeq 4\pi (4M^2)  (1-\frac 54 q)+O(q^2) \label{an}.
 \end{equation}

In the spherically symmetric case (\ref{an}) brings us back to the well known result
 \begin{equation}
 A_{Sch.}=16\pi M^2  \label{an1}.
 \end{equation}

\section{The surface gravity}

The surface gravity $k$ may be  defined from the time--translation Killing vector $\xi$ as
\begin{equation}
k^2\equiv-\frac12 \left(\nabla_{\mu}\xi_{\nu}\right)\left(\nabla^{\mu}\xi^{\nu}\right).
\end{equation}

In a static, asymptotically flat space-time it is the acceleration of a static observer near the horizon, as measured by a static observer at infinity.

Also it may be defined more generally for any surface outside the horizon in terms of the four--velocity of a fiducial observer at rest with respect to the frame   for which our metric functions are defined \cite{abreu}. At event horizon of course both definitions coincide.

For the case of a static and axisymmetric line element (diagonal matrix) the surface gravity $k$ is given by the following expression:
\begin{equation}
 k= \sqrt{-\frac{1}{4} g^{00} g^{ii}\left(\frac{\partial g_{00}}{\partial x^i}\right)^2},\label{surfaceg}
\end{equation}
where the index $i=1,2$  sums for the non-axial coordinates.

 By taking into account the expression (\ref{g00}) we have that:
\begin{eqnarray}
&\displaystyle{\partial_rg_{00}=2\left(-\frac{M}{r^2}-3\frac{Q}{r^4}P_2(\cos \theta)\right),\nonumber}\\
&\displaystyle{\partial_{\theta}g_{00}=-\frac{Q}{r^3} 6 \cos\theta \sin\theta},
\end{eqnarray}
and therefore the surface gravity for the M--Q solution at the event horizon is given by the following expression
\begin{equation}
 k=\frac{\hat\lambda^2 \sqrt{2}}{M \sqrt{1-2\hat\lambda-2\hat\lambda^3q P_2(y)}}\left[ \frac{(1-2
\hat\lambda)(1+3q\hat\lambda^3P_2(y))^2}{1+(1-2\hat\lambda)U}+\frac{9\hat\lambda^4q^2y^2(1-y^2)}{1+D}\right]^{1/2},\label{lak}
\end{equation}
where $\hat\lambda\equiv M/r$, and $U$, $D$ denote the series appearing at the metric functions $g_{rr}$ and $g_{yy}$ respectively which  are given by expressions (\ref{chachis}).

In the spherical case  we have $q=0$, $D=0$ and $U=0$  producing
\begin{equation}
k=\frac{\hat\lambda^2}{M},
\label{kschwar}
\end{equation}
which at the event horizon $r=2M$, ($\hat\lambda=1/2$) leads to the well-known expression for the surface gravity of Schwarzschild space-time:
\begin{equation}
k_{Sch.}=\frac{1}{4M} .
\end{equation}

Let us now calculate the surface gravity for the  M--Q solution  by considering the event horizon $r(y)$ as an small deviation with respect to the Schwarzschild surface $r=2M$ as follows:  $\displaystyle{\frac rM=2+p}$ where $p\equiv \epsilon/M$, and $\displaystyle{\hat\lambda\simeq \frac 12(1-\frac p2)+O(p^2)}$. Hence, the expression (\ref{lak}) just at order zero in power series  of the parameter $p$ leads to a divergent term at $y=y^0_1=1/\sqrt{3}$ since the surface gravity is given by
\begin{equation}
k\simeq \frac{3}{4} y \sqrt{g^{yy}_{(\hat\lambda=1/2)}}\sqrt{\frac{2q}{1-3y^2}}+O(p).
\end{equation}

An analogous conclusion follows for   the M--Q$^{(1)}$ solution. Indeed, in this case we have from (\ref{mq1})-(\ref{mq1detal})
\begin{eqnarray}
 g_{00}&=& -e^{2\Psi},\nonumber\\
g_{xx}&=&M^2e^{-2\Psi}e^{2\gamma}\frac{x^2-y^2}{x^2-1},\nonumber\\
g_{yy}&=&M^2e^{-2\Psi}e^{2\gamma}\frac{x^2-y^2}{1-y^2},
\end{eqnarray}
from which it follows at once,

\begin{equation}
 g^{00} g^{ii}\left(\frac{\partial g_{00}}{\partial
x^i}\right)^2=-\frac{e^{4\Psi-2\gamma}}{M^2(x^2-y^2)}\left[(x^2-1)X^2+(1-y^2)Y^2\right],\label{XY}
\end{equation}
where the following notation has been introduced
\begin{eqnarray}
 X&=&A_x \ln\left(\frac{x-1}{x+1}\right)-B_x+\frac{2A}{x^2-1},\nonumber\\
Y&=&A_y\ln\left(\frac{x-1}{x+1}\right)-B_y.
\end{eqnarray}
In  the spherical case  ($q=0$), as we approach the event horizon  ($x\rightarrow 1$) we have:
\begin{equation}
X=\frac{2}{x^2-1} \quad , \quad Y=0 ,
\end{equation}
producing with (\ref{surfaceg})
\begin{equation}
 k_{Sch.}=\frac{1}{4M}.
\end{equation}

If we now consider the  parameter $q$ to be arbitrarily small but non--vanishing, then from (\ref{XY}) we obtain
\begin{eqnarray}
 X^2&\simeq&\frac{4}{(x^2-1)^2}+\frac{15}{x^2-1}q\left[\frac{3y^2-1}{x^2-1}+\frac{x
(3y^2-1)}{2}\ln\left(\frac{x-1}{x+1}\right)+\right.\nonumber\\
&+&\left. \frac 16\left(-\frac{4}{1-y^2}+\frac{8}{(1-y^2)^2}+3(3y^2-1)\right)\right]+O(q^2),\nonumber\\
Y^2&\simeq&O(q^2),
\end{eqnarray}
and therefore the surface gravity for the M--Q$^{(1)}$ solution is given by the following expression
\begin{equation}
 k\simeq \frac{1}{4M}\left[1+\frac 58 q
(1-15y^2)+\Lambda\right]+O(q^2),
\end{equation}
where
\begin{equation}
\Lambda\equiv \frac{5}{2}q\lim_{x\rightarrow 1}(x-1)\left[\frac{y^2+1}{(1-y^2)^2}\right].\label{limit}
\end{equation}
The important point is that  everywhere outside the symmetry axis ($y\neq\pm1$), $\Lambda$ vanishes (at the event horizon), whereas   for the points on the symmetry axis $y=\pm 1$ at the horizon,  $\Lambda$ goes to infinity thereby producing a divergent surface gravity. This particular behaviour  exhibited in the limit of   (\ref{limit}) is characteristic of the M--Q space-time, and has been brought out  before in \cite{herreramq}.

It should be stressed that this divergence of the surface gravity on the event horizon at  $y=\pm 1$,  $\Lambda$ is not a ``directional'' singularity, i.e. surface gravity would diverge at  the points  $y=\pm 1$,  on the horizon, no matter how we approach it, as can be seen from (\ref{limit}). Therefore  such  a divergence is not related  to a breakdown of the coordinate system as seems to be the case for ``truly''  directional singularities \cite{taylor}.

Finally, it should be observed  that consistency with  third law of thermodynamics  would require surface gravity to go to zero when the area goes to zero. This condition is not satisfied for the gamma metric (see discussion in \cite{ita}). In our case however the area of the horizon remains finite.

We shall next discuss about the consequences emerging from the results presented so far.

\section{Conclusions}

We have presented a detailed description of the event horizon for the M--Q solution. The range of values of $q$ for which that surface is regular, closed, continuous and differentiable has been clearly established.

We have next calculated the area of  the event horizon. This quantity plays a central role in the definition of different bounds for black hole entropy (see \cite{abreu}, \cite{H4}--\cite{S} and references therein).

Now, if the gravitational  field of the collapsing body,  including  small deviations from spherical symmetry,  is described by means of  an exact solution to Einstein field equations (instead of perturbations of Schwarzschild space-time) then it  appears that  in some particular cases (e.g. the $\gamma$ space-time) the area surface of the event horizon vanishes leading to the conclusion that   as the body falls through the horizon, information about the collapsing body is stripped away,  thereby resolving the information loss paradox \cite{zipoVII}. However as we have seen before this is not the case for the M--Q solution, therefore the conclusion in \cite{zipoVII} seems to lacks of  universality, and does not apply to any deviation from the spherically symmetric case. Thus all the discussion about entropy bounds for spherically symmetric black holes remains basically unchanged  for slightly distorted space-times (small $q$) in the case of the M--Q solution.

On the other hand however, the result obtained for the surface gravity indicating that it diverges, points in the same direction that the one obtained for the $\gamma$ metric \cite{ita} (nevertheless observe that in our  case $k$ only diverges on the axis of symmetry whereas  in the $\gamma$ metric case it diverges everywhere except on the symmetry axis).

Now, the consequences of such divergence might be important and deserves a deeper analysis.

On the one hand,  the divergence of $k$ would imply that the ``inertial'' term in the transport equation would grow unlimited as the horizon is approached.

Indeed, as shown by Tolman many years ago \cite{Tolman}   according to special relativity all forms of energy have inertia, and of course this should also apply to heat. Therefore, because of the equivalence principle, there should be also some weight associated to heat, and one should expect that thermal energy tends to displace to regions of lower gravitational potential. This in turn implies that the condition of thermal equilibrium in the presence of a gravitational field must change with respect to its form in  absence of gravity. Thus a temperature gradient is necessary in thermal equilibrium in order to prevent the flow of heat from regions of higher to lower gravitational potential. This result was confirmed some years later by Eckart \cite{Eckart} and Landau and Lifshitz \cite{landau}. In the transport equation derived by these authors, the ``inertial'' term deduced by Tolman appears explicitly and is essentially the surface gravity multiplied by the temperature. It should be observed that exactly the same term appears also in transport equations derived from, more physically reasonable, causal thermodynamic theories (see \cite{TCR} for details).
In the case of Eckart and Landau and Lifshitz approach, the transport equation reads
\begin{equation}
q^{\alpha}=-\kappa h^{\alpha\beta}(T_{,\beta}+Ta_{\beta}),
\label{21}
\end{equation}
where $q^\mu$ is the heat flow four--vector,  $V^\mu$ is the four--velocity of a fiducial observer  mentioned above, $a^\mu$ is the four--acceleration associated to those fiducial observers, $h^{\mu \nu }$ is the projector onto the three space orthogonal to $V^\mu$, $\kappa$  denotes the thermal conductivity, and  $T$ denotes temperature.

Therefore the unlimited growing of  the ``inertial'' term as the surface of the object is closer to the event horizon would lead to an increasing inwardly directed heat flow vector, which in turn would produce a substantial increase of temperature.
Now, it has been shown (see \cite{TCR}, \cite{TC} and references therein) that, as the system leaves the equilibrium, the inertial mass  decreases by a factor  $(1-\alpha)$ where $\alpha$ is defined by
\begin{equation}\alpha=\frac{\kappa T}{\tau (\mu+P_r)}.\label{alpha}\end{equation}
where $\tau$ is the thermal relaxation time  and $\mu$ and $P_r$ denote the energy density and the radial pressure of the fluid respectively.

Obviously for this effect to be of some relevance in the dynamics of the system, evolution should proceeds in such a way that $\alpha$ approaches the critical value of $1$.

Now in c.g.s. units (omitting the pressure term which is always smaller than the energy density) we have
 \begin{equation}
\alpha\equiv \frac{\kappa T}{\tau \mu} \approx \frac{1}{81} \,
\frac{[\kappa]\,[T]}{[\tau]\,[\mu]} \times 10^{-40}.
\label{men40}
\end{equation}
where $[\kappa]$, $[T]$, $[\tau]$ and  $[\mu]$ denote
the numerical value of these quantities in $erg \, s^{-1} \, cm^{-1}
\, K^{-1}$, $K$, $s$ and $g \, cm^{-3}$ respectively.

Thus in order  for $\alpha$ to attain values close to the unity we need extremely high values of thermal conductivity and/or temperature. In the past \cite{Ma} it has been suggested that neutrino radiative heat conduction might produce values of $[\kappa]$ as large as $10^{37}$, which together with the expected values of temperature $[T]\approx 10^{13}$ at the last stages of massive star evolution, would lead to $\alpha\approx 1$.

Here we see that perturbations of spherical symmetry  described  by means of  the M--Q space-time (this also applies to the $\gamma$ metric case), could provide a mechanism for attaining large values of temperature due to the increasing in the heat flow  as the source approach the horizon.

On the other hand, starting from the generalized version of the surface gravity, it appears that its divergence as we approach the horizon, implies the divergence of the Tolman mass (see \cite{abreu}). This result whose dynamical implications cannot be overemphasize, is also obtained for the $\gamma$ metric (see \cite{mtlae} for a detailed discussion on this point).

Before concluding, the following  comment is in order:

According to the black hole has no--hair theorem, in the process of
contraction all (radiatable) multipole moments are radiated away \cite{Price}. Therefore it could be asked what the interest  might be to study singular horizons as the one corresponding to the M--Q solution?

Nevertheless, the situation is more complex than it looks at first sight. Indeed, let us admit that in the process of collapse, all (radiatable) multipole moments are radiated away. Obviously, the mechanism of radiation, as any physical process, must act at some time scale (say $\tau_{mech.}$). Now,  if  $\tau_{mech.}$ is
smaller than the time scale for any physical process occurring on the object (say
$\tau_{phys.}$), then the appearance of a regular  horizon  proceeds safely.

However, let us suppose for a moment that there is a physical process whose  $\tau_{phys.}$
is of the order of magnitude of (or still worse, smaller than) $\tau_{mech.}$. In this case any physical experiment based on such process ``will see'' a
singularity as  the boundary of the object crosses the horizon, due to the always present fluctuations.

Indeed, the fact remains that  perturbations of spherical symmetry take place all along the evolution of the object. Thus, even if it is true that close to the horizon, any of these
perturbations is radiated away, it is likewise true that this is a continuous process. Then, as soon as a  ``hair'' is radiated away, a new  perturbation appears which will be later
radiated and so on. Therefore, since  ``hairs'' are radiated away at some {\it finite} time scale, then at that time scale ($\tau_{mech.}$) there will be always a  fluctuation acting
on the system.

Thus, unless one can  prove that indeed  $\tau_{mech.}$ is smaller than
$\tau_{phys.}$ {\it{for any physical process}}, one should  take into account the possible consequences derived from the presence of fluctuations of spherical symmetry (close to
the horizon), in this later case such deviations from spherical symmetry might be described (at least in some cases) by the M--Q space-time.

\section{Acknowledgments}
This  work  was partially supported by the Spanish  Ministerio de Ciencia e Innovaci\'on
 under Research Project No. FIS 2009-07238, and the Consejer\'\i a de Educaci\'on of the Junta de Castilla y
Le\'on under the Research Project Grupo de Excelencia GR234, and we also wish to thank the support of the Fundaci\'on Samuel Sol\'orzano Barruso (University of Salamanca) with the project FS/8-2010.

\begin{figure}[t]
\begin{minipage}{.53\linewidth}
\centering
\includegraphics[height=3.2in,width=3.2in]{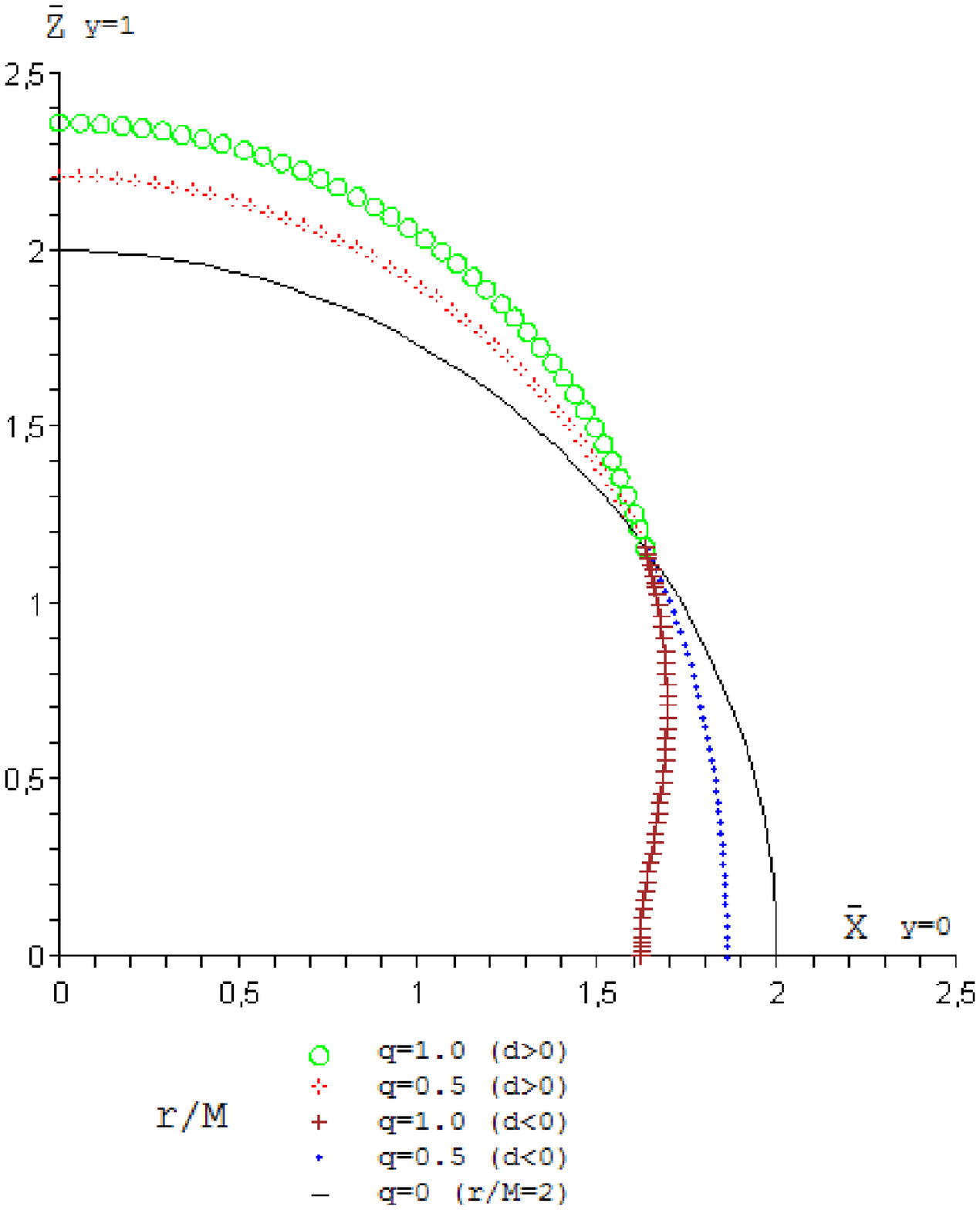}
{$q \geq 0$}
\end{minipage}
\begin{minipage}{.49\linewidth}
\centering
\includegraphics[height=3.2in,width=3.2in]{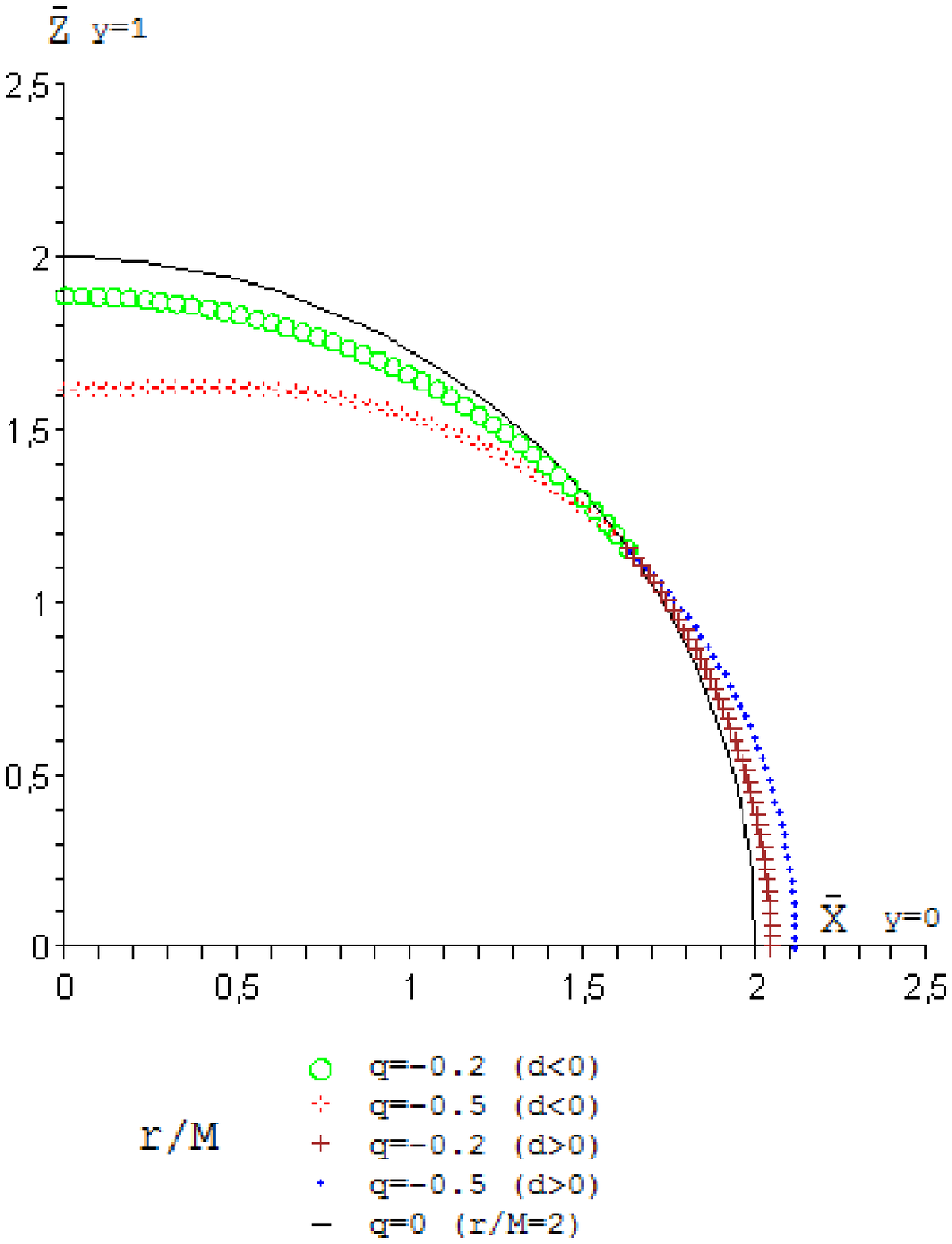}
{$q \leq 0$}
\end{minipage}
\caption{Generatrix curve $r=r(y)$, except for a factor $1/M$, in polar coordinates, for different positive values of $q$. The vertical axis $\bar Z$ represents  the symmetry axis $y=\pm 1$, whereas the horizontal abscissa axis $\bar X$ represents  anyone direction on the equatorial plane orthogonal to $\bar Z$. The curve for each value of $q$ is drawn in different color depending on the corresponding domain up or down the boundary value $y=y_1^0$, where all different curves are crossing themselves.}
\label{fig2}
\end{figure}

\begin{figure}[h]
\begin{minipage}{.49\linewidth}
\centering
\includegraphics[height=3.0in,width=2.2in]{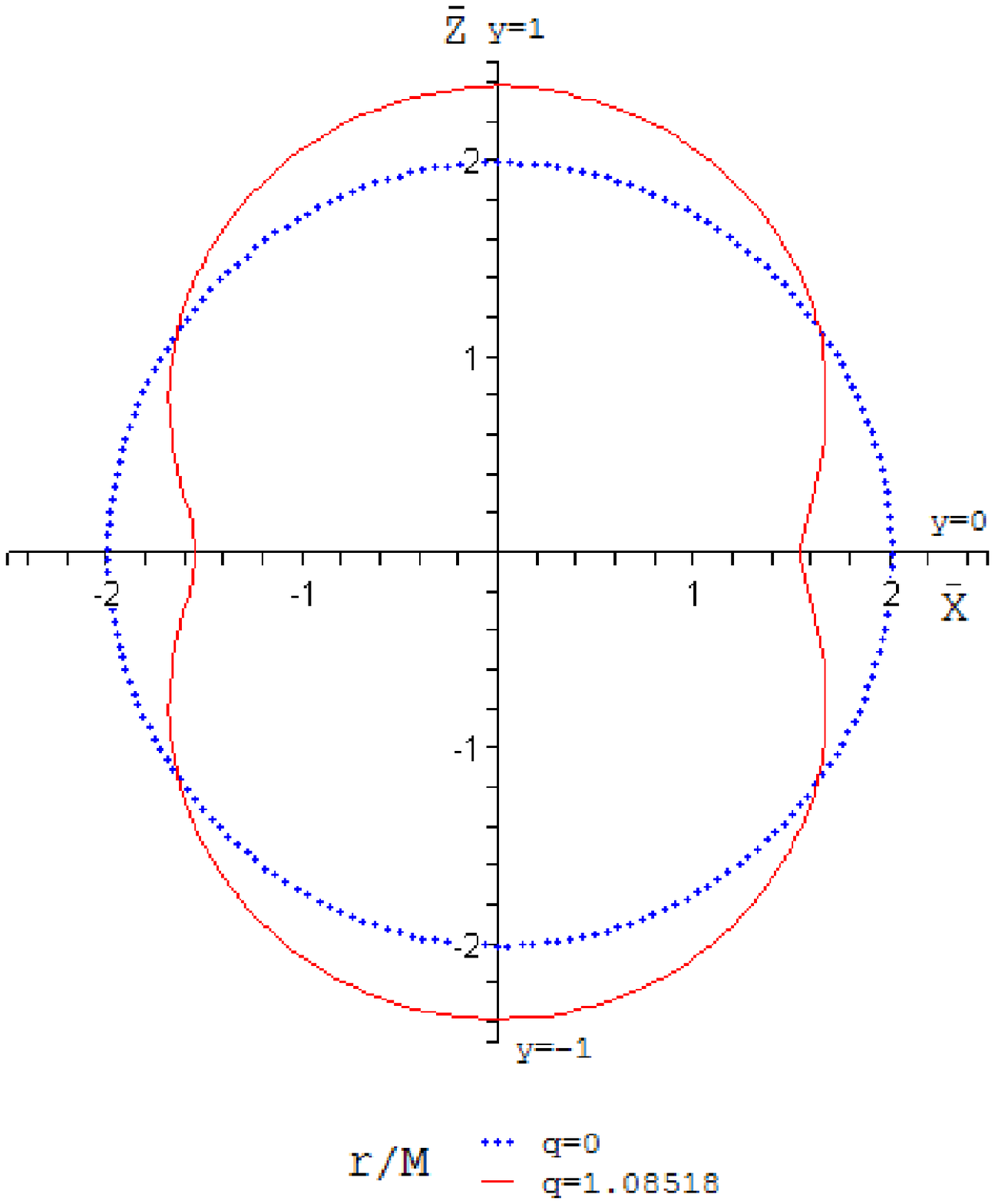}
%{}
\end{minipage}
\begin{minipage}{.49\linewidth}
\centering
\includegraphics[height=3.0in,width=2.2in]{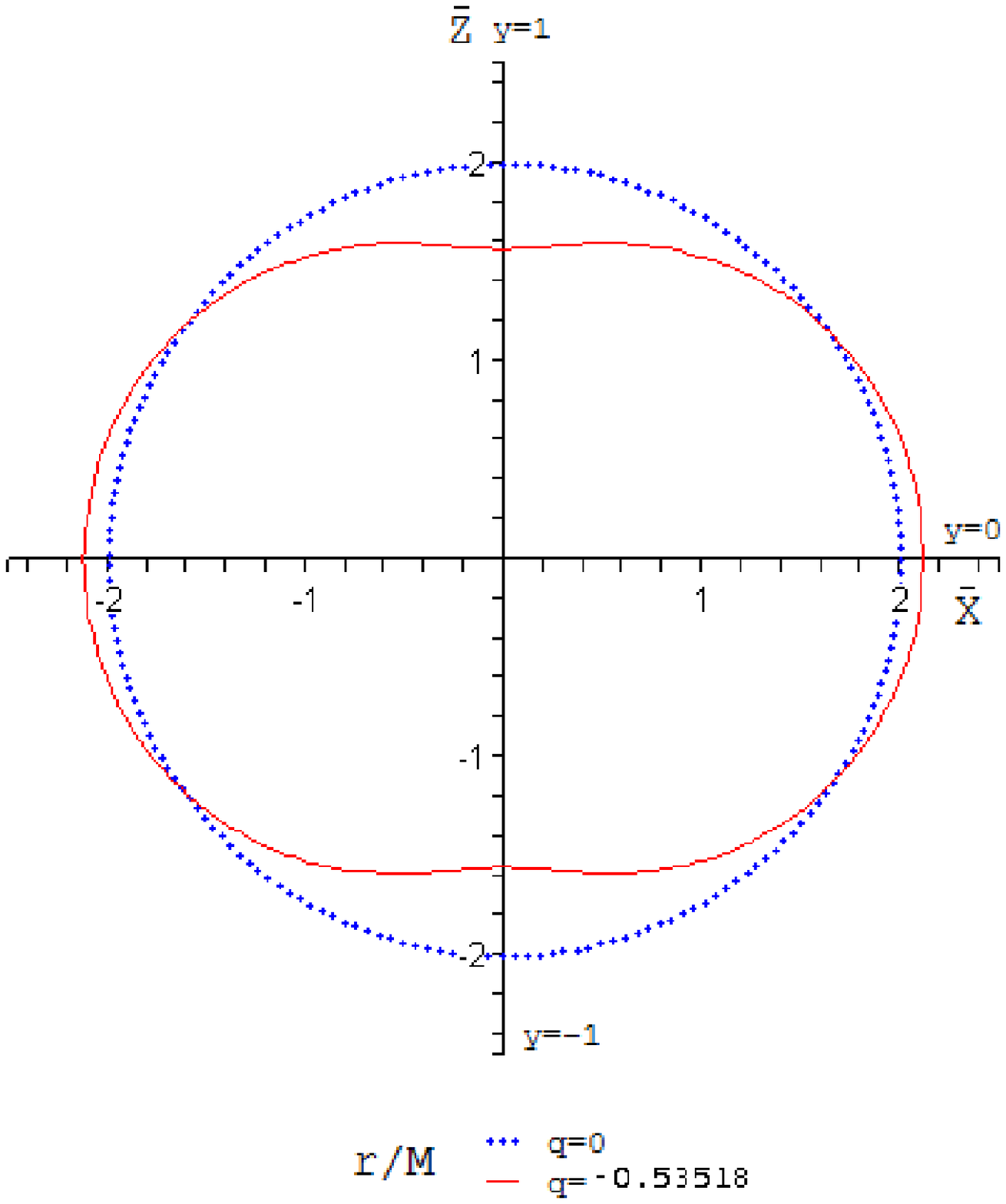}
%{}
\end{minipage}
\caption{The whole curve $r(y)/M$ for all values of the angular polar coordinate and different values of $q$ ($q=0$, $q=1.08518$ and $q=-0.53518$). The curve with points represents the event horizon for the spherical symmetry case.}
\label{fig3}
\end{figure}

\begin{figure}[h]
\begin{minipage}{.49\linewidth}
\centering
\includegraphics[height=2.5in,width=2.5in]{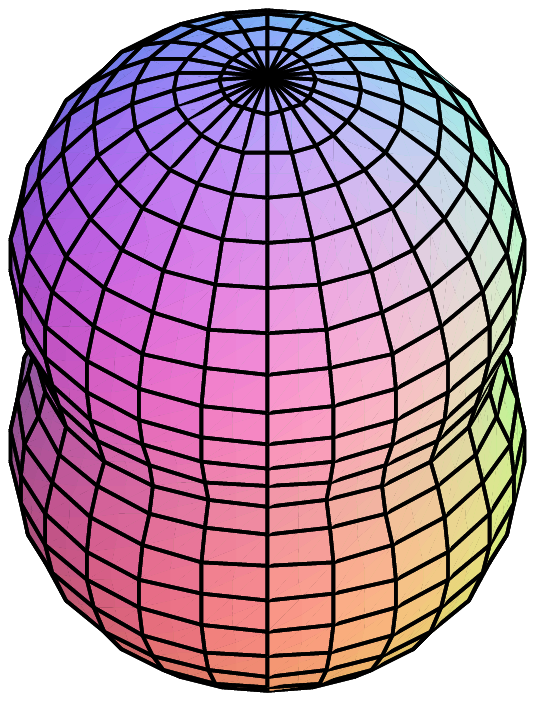}
{$q=1.12$}
\end{minipage}
\begin{minipage}{.49\linewidth}
\centering
\includegraphics[height=2.3in,width=2.3in]{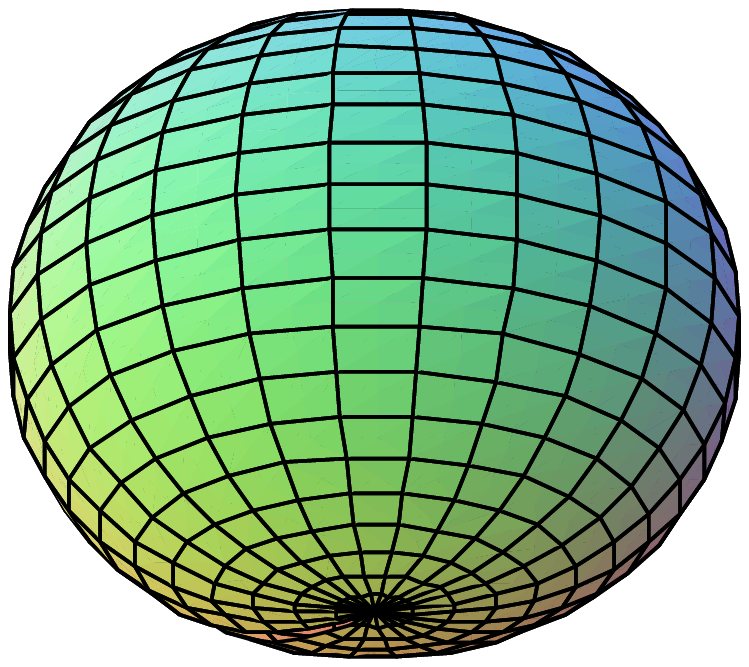}
{$q=-0.4$}
\end{minipage}
\caption{The surface of the event horizon for  different values of $q$.}
\label{fig4}
\end{figure}

\end{document}